\documentclass[iop]{emulateapj}
\usepackage{amssymb}
\usepackage{times}
\usepackage{epsfig}
\usepackage{graphics}

\newcommand{\be}{\begin{equation}}
\newcommand{\ee}{\end{equation}}
\newcommand{\bea}{\begin{eqnarray}}
\newcommand{\rmd}{{d}}
\newcommand{\eea}{\end{eqnarray}}

\usepackage[dvips]{color}
\usepackage[normalem]{ulem}
\usepackage{fancybox}

\slugcomment{ }
\shorttitle{Carbon NS atmospheres}
\shortauthors{Suleimanov, Klochkov, Pavlov, \& Werner}

\begin{document}

\title{ Carbon neutron star atmospheres
    }


\author{ V. F. Suleimanov\altaffilmark{1,2}, D. Klochkov \altaffilmark{1}, G.G. Pavlov\altaffilmark{3}, and K. Werner\altaffilmark{1}}
\affil{}
\email{}


\altaffiltext{1}{Institute for Astronomy and Astrophysics, Kepler Center for Astro and Particle Physics,
   Eberhard Karls University, Sand 1, 72076 T\"ubingen, Germany; suleimanov@astro.uni-tuebingen.de}
\altaffiltext{2}{Kazan (Volga region)
 Federal University, Kremlevskaya 18, 420008 Kazan, Russia}
\altaffiltext{3}{Pennsylvania State University, 525 Davey Lab., University Park, PA 16802;
pavlov@astro.psu.edu}
\begin{abstract}

The accuracy 
of measuring the basic parameters of neutron stars
is limited in particular by uncertainties in  chemical composition of their 
atmospheres.  For example,  atmospheres of thermally - emitting neutron stars in supernova 
remnants might have exotic chemical compositions, and for one of them, the neutron star in Cas~A, 
 a pure carbon atmosphere  has recently been suggested by \citet{HoH:09}.
To test such a composition for other similar sources, a publicly available detailed grid of 
carbon model atmosphere spectra is needed. We have computed such a grid
using the standard LTE approximation and assuming that the magnetic field does not exceed
  $10^8$\,G. The opacities and pressure ionization effects  are calculated  using the 
Opacity Project approach. We describe the properties of our models  and investigate the
impact of the adopted assumptions and approximations on the emergent spectra. 
\end{abstract}
\keywords{
radiative transfer ---methods: numerical -- stars: neutron -- stars: atmospheres -- X-rays: stars}

\section{Introduction}

Thanks to the outstanding capabilities of the {\sl Chandra} and {\sl XMM-Newton} X-ray
observatories, 
thermal radiation of a few classes of 
isolated (non-accreting) neutron stars (NSs)  
has been investigated 
\citep[see,
e.g.,][]{P:02a, WP:06}.  
Fitting the X-ray spectra of these objects with appropriate models,
one can measure the surface temperatures and radii of NSs.
Measuring the surface temperatues of NSs of different ages
allows one to study  the thermal evolution of NSs, which depends on
the properties of super-dense matter in their cores 
\citep[see, e.g.,][]{YP:04}. 
The NS radii provide 
important 
additional information about the super-dense matter 
as they are determined, for a given NS mass, by the equation of state,
which depends on the (currently unknown) composition of the NS core and
many-body interactions
 of the constituent particles 
(see, e.g., the review by
\citealt{LP:07}). 

Fitting the thermal spectrum of an isolated NS with a blackbody
model, one can determine the apparent blackbody temperature $T_{\rm BB}$
and the normalization parameter $K_{\rm BB}=R_{\rm BB}^2/d^2$, where
$R_{\rm BB}$ is the apparent blackbody radius and $d$ is the distance.
However, the actual thermal spectrum can differ significantly from the
Planck spectrum, and $T_{\rm BB}$ and $R_{\rm BB}$ can be different
from the actual temperature and radius. 
In particular, if the NS surface is
covered by an optically thick hydrogen or helium plasma envelope
(atmosphere), the emergent spectrum
is harder than the blackbody spectrum with the same effective temperature,
$B_{E}(T_{\rm eff})$,
 because photons with higher energies $E$ are emitted
from deeper, hotter layers due to the opacity decrease with increasing $E$
(e.g., the free-free opacity $k_{\rm ff}\propto E^{-3}$ at low 
magnetic fields) --- see, e.g., \citet{ZPS:96}.
Therefore, the temperature 
obtained from a blackbody fit is higher than the effective temperature, 
$T_{\rm BB} = f T_{\rm eff}$, where 
$f \approx 1.5$--3, while the blackbody radius is smaller than the true
radius because the observed flux does not depend on the model chosen for
fitting (e.g., $R_{\rm BB}\sim f^{-2} R$ if the observed energy range
includes a substantial fraction of the bolometric flux, $F_{\rm bol}=
(R^2/d^2)\sigma_{\rm SB}T_{\rm eff}^4$\footnote{We neglect General Relativity effects in these estimates 
(see Section 2).}. 
Thus, to measure the NS surface temperature and radius, one should
fit the observed spectra with model atmosphere spectra rather than with
the blackbody model.

Particularly useful 
 targets for measuring the NS temperatures and radii are the so-called
Central Compact Objects (CCOs) in shell-type supernova remnants.
These NSs show purely thermal spectra, uncontaminated by nonthermal emission,
and many of them have
relatively good distance estimates, which is important for measuring the radii
(see \citealt{P:02b}, \citealt{P:04}, \citealt{GHA:13} for reviews).
The apparent blackbody radii 
are about a few kilometers only, 
which is significantly smaller than the 10--15 km canonical NS radii.
 Some of these objects exhibit X-ray pulsations with a
 significant pulsed fraction, up to 64\% for the
CCO in SNR Kes 79 \citep{HG:10}. 
In such cases the small sizes of the emitting area can be explained by
the existence of relatively 
small hot spots on 
the NS surface.  In some other cases, fitting the spectra with pure
hydrogen atmosphere models yields
reasonable NS sizes. For example, the radius of the CCO 
1E\,1207.4$-$5209 obtained with a blackbody fit is  1--3 km,
while the radius derived using hydrogen atmosphere models is about 10
km \citep{ZPT:98}.  

An interesting and unusual case among
CCOs is the NS in the Cas\,A supernova remnant. 
Blackbody fits of first {\sl Chandra} observations of this source yielded
a high temperature and a small emitting radius
$T_{\rm BB}=6-8$\,MK, $R_{\rm BB} = 0.2-0.5$\,km, 
at a distance of 3.4 kpc \citep{P:00}
while the one-component blackbody fit of a
 {\sl Chandra} spectrum of better quality was statistically 
unacceptable \citep{PL:09}. The spectrum could be fit with a number 
of two-component thermal
models with substantially different temperatures and sizes
(e.g., $T_1\approx 4.5$ MK, $R_1\approx 0.4$ km, $T_2 \approx 1.6$ MK, $R_2\approx 12$ km 
for a two-component
hydrogen atmosphere model).
With such a small hot spot the source is expected
 to show  pulsations which, however, have not been detected
(the $3 \sigma$ upper limit on the pulsed fraction is about 16\%,
assuming a sinusoidal pulse shape; \citealt{PL:09}).
A possible solution to this problem has been
suggested by \citet{HoH:09}, who assumed that
the NS atmosphere 
has a chemical composition different from
pure hydrogen or helium 
and demonstrated that a pure carbon
atmosphere model spectrum gives 
a good fit 
to the data and a reasonable emitting area of 10--14\,km. 
Spectra of carbon atmospheres are  harder and have 
lower flux in the observed energy band 
 (0.5--6\,keV) than those of hydrogen
atmospheres  at a given $T_{\rm eff}$. However,
direct evidence for a carbon atmosphere 
is still missing, and the possibility of 
a NS with a hydrogen/helium hot spot on  
 its surface, pulsating with a low amplitude,  cannot be excluded.
 On the other hand,   
hydrogen and helium can diffuse from the atmosphere to deeper hot layers,
where they burn into carbon (Rosen 1968). 
As a result, a carbon atmosphere can form on the time scale of
a few years (Chang et al.\ 2010).
 
Recently a ``twin'' of the Cas\,A CCO has been discovered near the
center of the supernova remnant shell  
HESS~J1731$-$347 / G353.6$-$0.7 \citep{Ac:09}. 
The blackbody fit of its thermal X-ray spectrum 
yields $T_{\rm BB} \approx 6$ MK and 
$R_{\rm BB}
\approx 1$ km, for a distance of 3.2 kpc. An assumption of the canonical
NS radius of 10 km  
leads to the distance $>30$\,kpc, which is incompatible with the
properties of the supernova remnant \citep{Ab:11}. 
We have therefore suggested that
the NS in HESS~J1731$-$347 has a carbon atmosphere,
  similar to the Cas~A CCO \citep{Kl:13}.  
In the absence of 
publicly available carbon atmosphere models,
we have developed our own models and computed a large set of the model spectra. This
set was used to fit the {\sl XMM-Newton}
spectrum of the CCO and yielded a reasonable NS radius
$R\approx 13$ km and mass  
$M\approx 1.5 M_{\odot}$ 
\citep{Kl:13}. The same grid was used to fit 
two high-quality {\sl Chandra} spectra of the Cas~A CCO \citep{PPS13}.

In this paper, we provide a detailed description of our
carbon atmosphere model computation and 
describe properties of the computed models.
The model spectra
are available in form of an XSPEC model 
(model \texttt{carbatm}\footnote{http://heasarc.gsfc.nasa.gov/xanadu/xspec/models/carbatm.html}).
We also discuss possible uncertainties which might influence the
emergent spectra.

\section{Method of carbon atmosphere modeling}

The models presented here  are computed assuming  hydrostatic and
radiative equilibria in the plane-parallel  
approximation \citep[see][for details]{Mihalas:78}. 
The main input parameters are the chemical composition (we consider
here pure carbon atmospheres and atmospheres with carbon and hydrogen mix); 
the surface gravity 
\be \label{eq:g_def}
   g=\frac{GM}{R^2}(1+z)
\ee
(where $M$ and $R$ are the 
NS mass and radius, respectively),
and the effective temperature $T_{\rm eff}$.
The gravitational redshift $z$ on the stellar surface is related to the
NS mass and radius as follows 
\be \label{eq:redshift_def}
    1+z=(1-2GM/c^2R)^{-1/2} .
\ee

The structure of the atmosphere is described by a set of differential
equations. The first one is the hydrostatic equilibrium equation
\be \label{eq:hyd}
  \frac {d P_{\rm g}}{dm} = g - g_{\rm rad},
\ee
where $g_{\rm rad}$ is the radiative acceleration, 
$P_{\rm g}$  is the gas pressure, and the column density $m$ is
defined as
\be
\rmd m = -\rho \, \rmd s \, ,
\ee
where $s$ is the vertical distance and $\rho$ is the gas density.

The second equation is the radiative transfer equation for the
specific intensity $I_\nu (\mu)$.  
In the plane-parallel approximation, it has the form
\be \label{eq:rte}
\mu \frac{\rmd I_\nu (\mu)}{\rmd\tau _\nu } = I_\nu (\mu) - 
\frac{k_\nu}{\sigma_{\rm e} +k_\nu }B_\nu - \frac{\sigma_{\rm e}}{\sigma_{\rm e} +k_\nu }J_\nu,
\ee
where 
$\mu = \cos \theta$ is the cosine of the angle between the surface
normal  and the direction of radiation propagation,  
$k_\nu$ is  the ``true absorption'' opacity,
which includes bound-bound, bound-free and free-free transitions, 
$\sigma_{\rm e}$ is
the coherent Thomson  electron scattering opacity,
\be
     J_\nu =\frac{1}{2}\, \int^{+1}_{-1} \, I_\nu(\mu)\, \rmd\mu 
\ee
is the mean intensity (the zeroth moment of specific intensity), and $B_{\nu}$ is the Planck function. 
We consider relatively cold model atmospheres (the effective temperatures are 
less than 4\,MK)  and, therefore, we include only the coherent electron scattering.
Here we also neglect the electron scattering anisotropy and
 radiation polarization, 
which is  
usual in the stellar atmosphere modeling 
because the emergent flux errors caused by this approximation are negligibly small 
(see, e.g., Chandrasekhar 1946).

The radiation pressure acceleration $g_{\rm rad}$ 
can be calculated as
\be \label{eq:grad_stan}
g_{\rm rad} = \frac{4\pi}{c} \, 
\int^{\infty}_{0}   \left(\sigma_{\rm e} +k_\nu\right)\,   H_\nu \ \rmd \nu , 
\ee
where
\be
     H_\nu =\frac{1}{2}\, \int^{+1}_{-1} \mu\, I_\nu(\mu)\, \rmd\mu 
\ee
is the Eddington flux (the first moment of specific intensity).

The set of equations is completed  by the energy balance equation
\be  \label{eq:econs}
\int^{\infty}_{0}  k_\nu\, \left(J_\nu-B_\nu\right) \, \, d\nu = 0,
\ee
the ideal gas law
\be   \label{gstat}
    P_{\rm g} = N_{\rm tot}\ kT,
\ee
where $N_{\rm tot}$ is the number density of all particles, and  the
particle number and charge conservation equations.  
Although the equation of state can deviate from the ideal gas law 
at very high densities and/or low temperatures
(see, e.g.,  Hummer \& Mihalas 1988; Rogers \& Iglesias 1992; Potekhin 1996),
this deviation should not affect our atmosphere models substantially,
as we discuss  in the next Section.

We assume local thermodynamic equilibrium
(LTE) and calculate the number densities of all ionization and excitation states 
using the Boltzmann and Saha equations. 
We accounted for  the pressure ionization and level dissolution effects  
using the occupation probability formalism \citep{Hum.Mih:88} as  described by \citet{Lanz.Hub:94}.
We use 96 levels of the CVI ion and 
11 levels of the CV ion 
to calculate the partition functions
 of these most abundant bound ions.    
In addition to the electron scattering,  we take into account the free-free opacity 
as well as the bound-free transitions  
for all carbon ions.  The bound-free opacities due to photoionization from 
the ground states of all ions 
were computed using the routine presented in  \citet{VYa:95} and \citet{V:96} \citep[see][]{Ibragimov.etal:03}.
In addition, photoionization from 
5 excited levels of the CVI ion 
and from 
10 excited levels of the CV ion are
included in the opacity  
using the approach
by \cite{Karz.L:61}  and the corresponding Kurucz's subroutine.
We note that 
 such calculations are rather approximate for 
the CV ion
because they replace the C nucleus and the inner (non-excited) electron by
a pointlike Coulomb potential,  
but their accuracy is sufficient for 
the considered atmosphere models
because the number densities of the excited CV ions are 
small.
 The bremsstrahlung
opacities of all the ions 
are calculated 
assuming that the ion's electric field 
is a Coulomb field of a charge $Ze$ equal to the ion's charge and using the
Gaunt factors from Sutherland (1998).
 
Line blanketing is taken into account using carbon spectral lines from the atomic database CHIANTI, Version 3.0
\citep{dere:97}. The Stark broadening of the H-like CVI
ion lines is considered 
following \citet{Griem:60} while for
the other lines it is treated
according to \citet{Cow:71}. 

To solve the above equations, we
used a version of the computer code ATLAS \citep{Kurucz:70,Kurucz:93},
modified to deal with high temperatures \citep[see][]{Ibragimov.etal:03,sw:07,RSW08} with the updates 
described above.  The radiative transfer equation is solved using the short characteristic method
\citep{OK87}. We used $\sim$$20\,000$ logarithmically equidistant
frequency points in 
the range $10^{13}$--$5\times 10^{19}$ Hz  ($\sim$10$^{-5}$--40 keV) 
for an accurate treatment of the line blanketing (by the opacity
sampling method). The calculations are performed for a set of 98
depth points $m_{\rm i}$ distributed equidistantly in a logarithmic
scale from $\sim$10$^{-6}$ to $m_{\rm max} \sim 10^5$~g~cm$^{-2}$. 
The appropriate value of $m_{\rm max}$ is chosen to satisfy the
condition $\tau^{\rm eff}_{\nu}(m_{\rm max}) > 1$ at all frequencies,
 where $\tau^{\rm eff}_\nu$ is determined by the equation
$$
     d\tau_\nu^{\rm eff} = \sqrt{k_\nu (k_\nu+\sigma_{\rm e})}\, \,dm.
$$ 
This requirement is necessary to satisfy the
inner boundary condition of the radiative transfer problem (the
diffusion approximation). 

\begin{figure}
\centering
\includegraphics[angle=0,scale=1.]{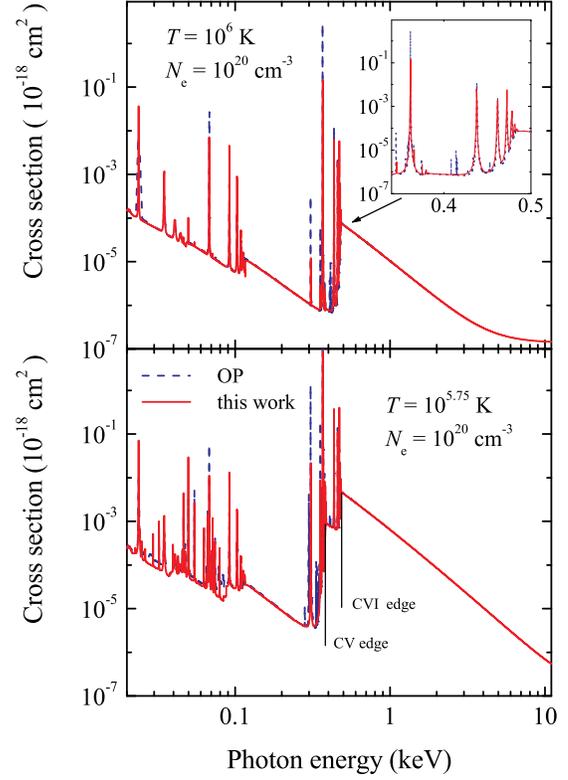}
\caption{\label{af0}
Comparison of the carbon 
opacities published by Opacity Project (dashed curves) and computed  in this work (solid 
curves) for two 
temperatures. 
The inset graph in the
  top panel shows a zoom into the region of the Lyman-like lines for
  the higher temperature.
The  ionization threshold
energies of CVI and CV ions, 489 and 392 eV, respectively, are indicated
by solid vertical lines in the bottom panel.   
}
\end{figure}

We compared the opacity calculated in this work with the carbon
opacities presented by the Opacity
Project\footnote{http://cdsweb.u-strasbg.fr/topbase/testop/TheOP.html}
\citep[OP;][]{S94}. 
The comparison has been done for two temperatures, 
$T = 10^6$~K and $T = 10^{5.75}$~K, 
at the electron number density $N_e = 10^{20}$\,cm$^{-3}$.
Figure \ref{af0}
shows that our approach 
is sufficiently accurate, including the width of the  
H-like spectral lines. In particular, the comparison demonstrates that
our calculations of the number
densities of both CV and CVI ions 
and the populations of their excited states 
agree with the OP results. 

\section{Basic properties of the models}

\begin{figure}
\centering
\includegraphics[angle=0,scale=0.8]{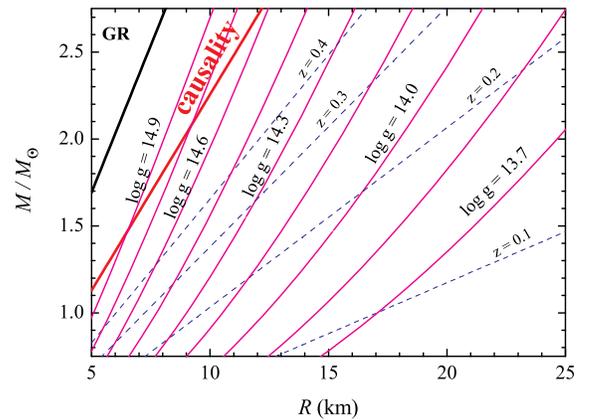}
\caption{\label{af0a}
Lines of constant $\log g$ values, for which our models were computed,
 in the $M$--$R$ plane. The  limits of the forbidden 
regions in the upper-left corner of the plane are 
shown by thick lines.
The dashed lines are the lines of constant
gravitational redshift $z$.
}
\end{figure}

Using the code described above, we have calculated an extended set  of
 carbon NS atmosphere models. The
models are computed for 
9 values 
of surface gravity $\log g$, from 13.7 to 14.9 with a step of 0.15 ,
which cover most of realistic  
NS equations of state for a wide range of NS masses (Fig.\,\ref{af0a}). 
For every value
of $\log g$, we computed 61 models with effective temperatures  
$T_{\rm eff}$ from 1 to 4\,MK with a step of
0.05\,MK.

\begin{figure}
\centering
\includegraphics[angle=0,scale=1.]{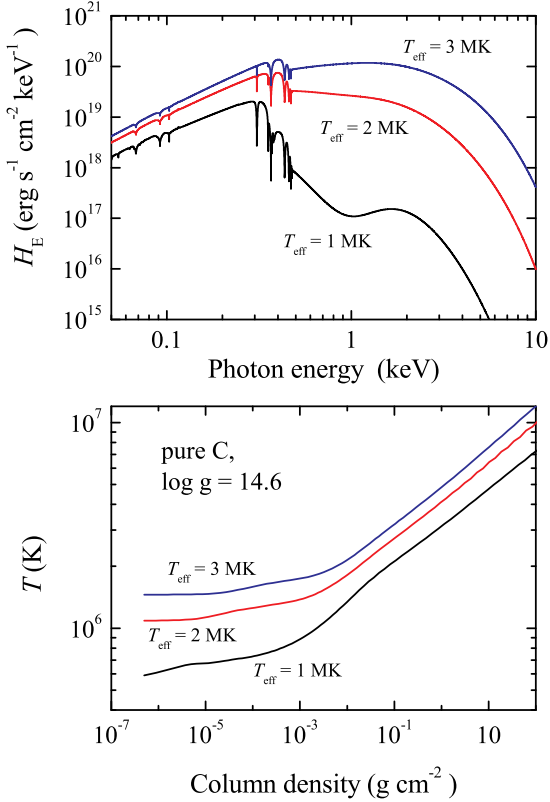}
\caption{\label{af1}
Emergent spectra (top) and temperature structures (bottom)
 of pure carbon atmosphere models with 
$\log g = 14.6$
and different effective temperatures.
}
\end{figure}

\begin{figure}
\centering
\includegraphics[angle=0,scale=1.]{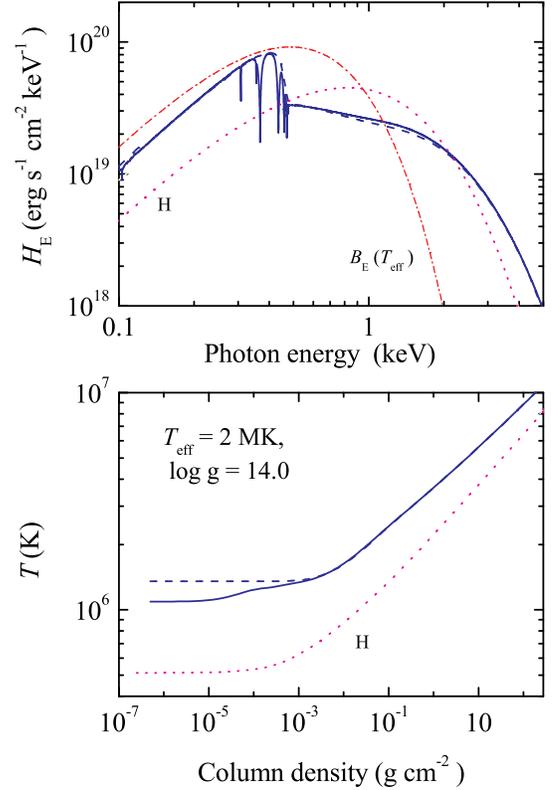}
\caption{\label{af2}
Comparison of the emergent spectra (top) and temperature structures (bottom) of 
pure carbon atmosphere models computed with and without
spectral lines (solid and dashed curves, respectively). The spectrum and  
the temperature structure of the pure hydrogen model are also shown (dotted curves). 
All the models have $T_{\rm eff}$ = 2\,MK and $\log g$ = 14.0.
The Planck function corresponding to 2\,MK is shown in the top panel
by the dash-dotted curve. 
}
\end{figure}

\begin{figure}
\centering
\includegraphics[angle=0,scale=1.]{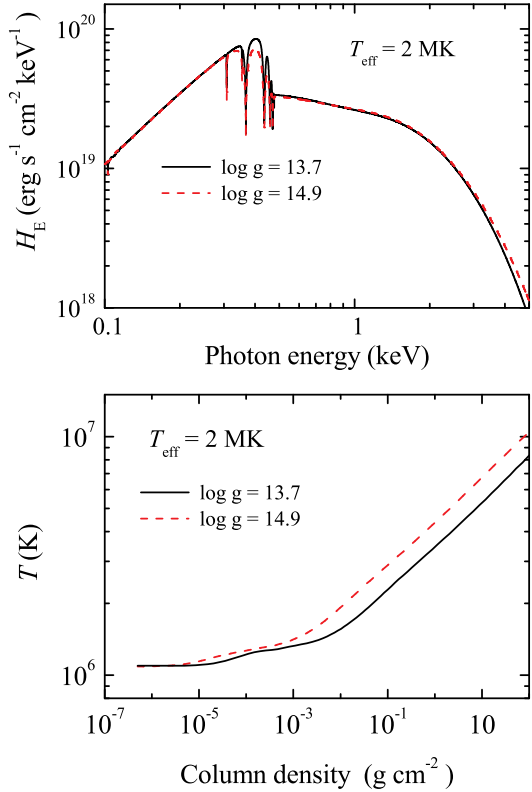}
\caption{\label{af3}
Emergent spectra (top) and temperature structures (bottom) 
of pure carbon atmosphere models with 
$T_{\rm eff}$ = 2 MK and two extreme values of $\log g$, 14.9 and 13.7.
}
\end{figure}

\begin{figure}
\centering
\includegraphics[angle=0,scale=1.]{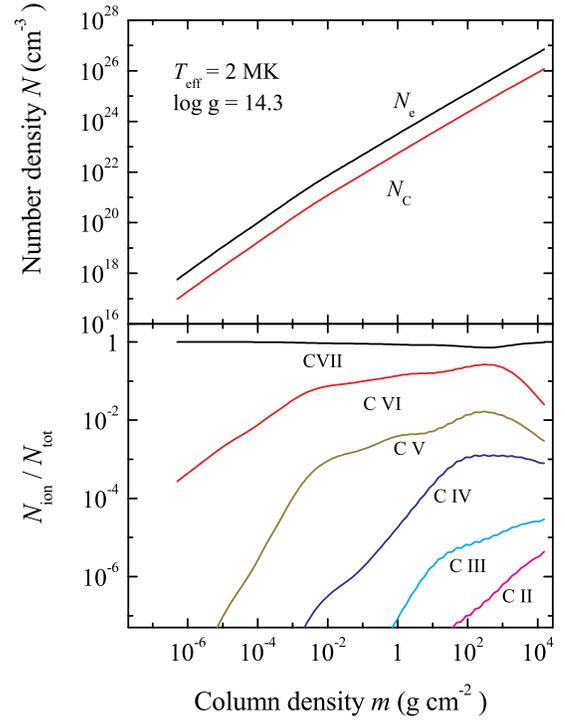}
\caption{\label{dens-ioniz} 
Dependence of the number densities of electrons, $N_e$, and carbon ions, $N_C$
(top), and the ionization fractions (bottom) on the column depth for the 
carbon atmosphere model with $T_{\rm eff}=2$ MK and $\log g = 14.3$.
}
\end{figure}

\begin{figure}
\centering
\includegraphics[angle=0,scale=1.]{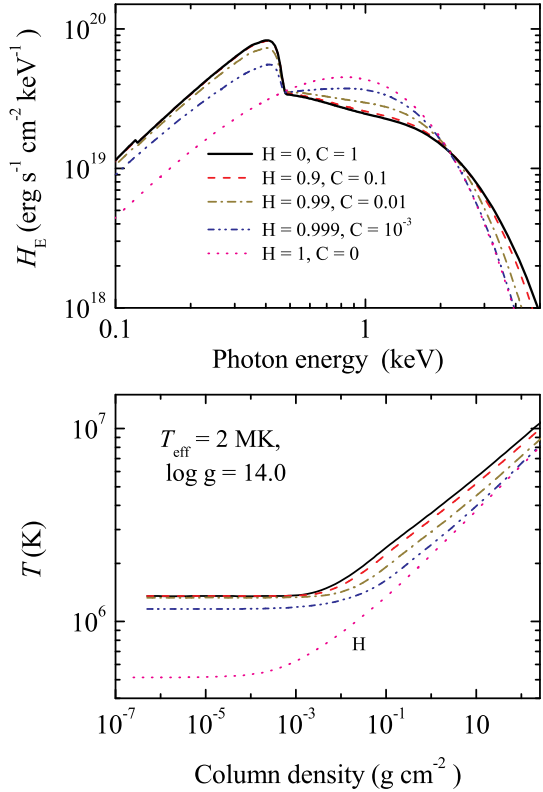}
\caption{\label{af4}
Comparison of the emergent spectra (top) and temperature structures
(bottom) for hydrogen+carbon models with 
different hydrogen/carbon fractions, from pure
carbon to pure hydrogen.  All the models correspond to
$T_{\rm eff}$ = 2\,MK and $\log g$ = 14.0 and were computed without taking
into account spectral lines.
}
\end{figure}

\begin{figure}
\centering
\includegraphics[angle=0,scale=1.]{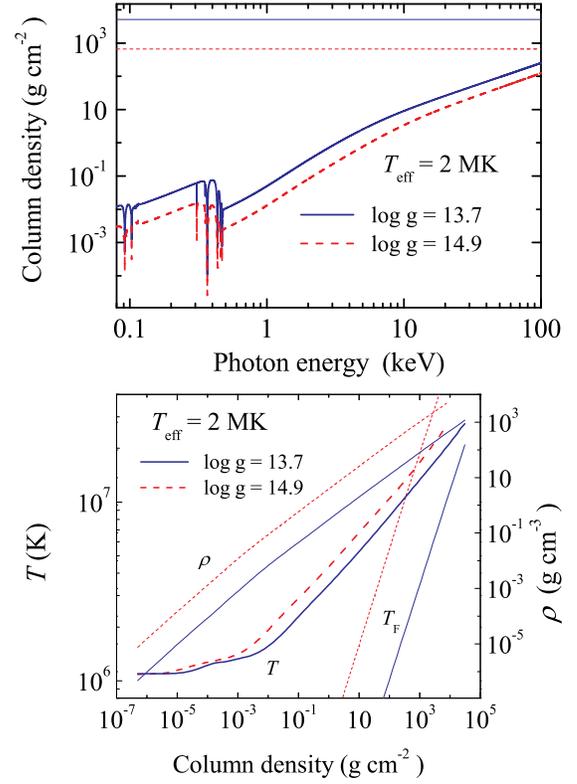}
\caption{\label{af5}
\emph{Top:} Comparison of the spectrum formation depths
($\tau_\nu^{\rm eff}$ = 1) with the depths 
 where the Fermi temperature becomes larger than the model gas
 temperature (horizontal lines) for two pure carbon models: 
 with $\log g$ = 13.7 (solid curves)
 and $\log g$ = 14.9 (dashed  
 curves). Both models correspond to $T_{\rm eff}$ = 2\,MK.
 \emph{Bottom:} Temperature structures and the corresponding Fermi temperature distributions
together with density distributions  for the same
models.
}
\end{figure}

\begin{figure}
\centering
\includegraphics[angle=0,scale=1.]{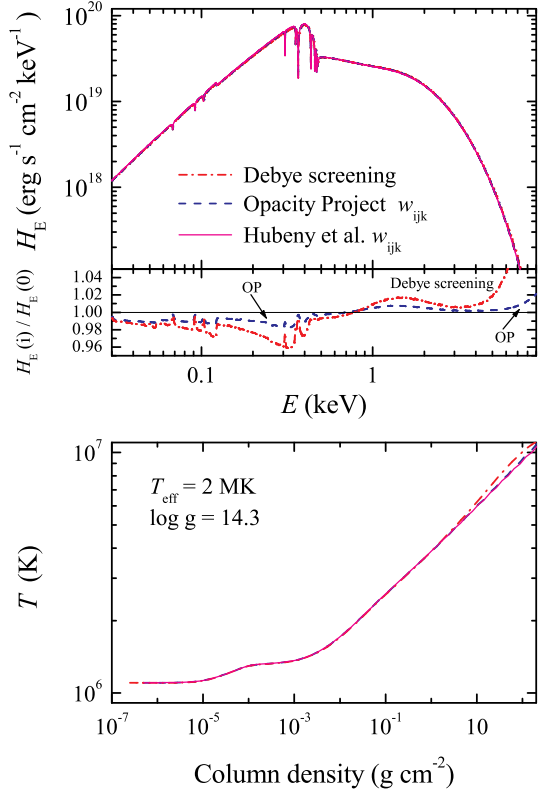}
\caption{\label{af5a}
Comparison of the emergent spectra (top) and temperature structures (bottom)
 of pure carbon atmospheres computed using three different approaches to
account for pressure ionization (see text). 
The solid curves correspond to the computations using the
  approximation of \cite{Lanz.Hub:94} adopted in our work,
the dash-dotted curves 
correspond to the computations using a simple
  Debye screening, and the dashed curves indicate the models computed
  using the Opacity Project approximation.
The ratios of the spectra of two latter models to the spectrum of the 
 first one are shown in the sub-plot
of the top panel.
}
\end{figure}

\begin{figure}
\centering
\includegraphics[angle=0,scale=1.]{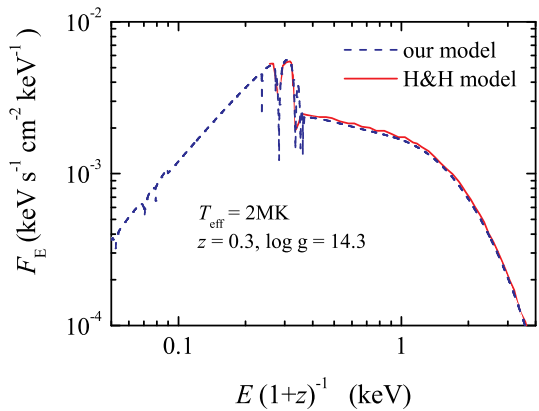}
\caption{\label{af6}
Comparison of our 
redshifted model spectrum ($\log g$ = 14.3, 
$T_{\rm eff}$ = 2\,MK, $z=0.3$,  $d=3.4$\,kpc, dashed curve),
 with the \citet{HoH:09} 
model spectrum (solid curve).
}
\end{figure}
 
\begin{figure}
\centering
\includegraphics[angle=0,scale=0.5]{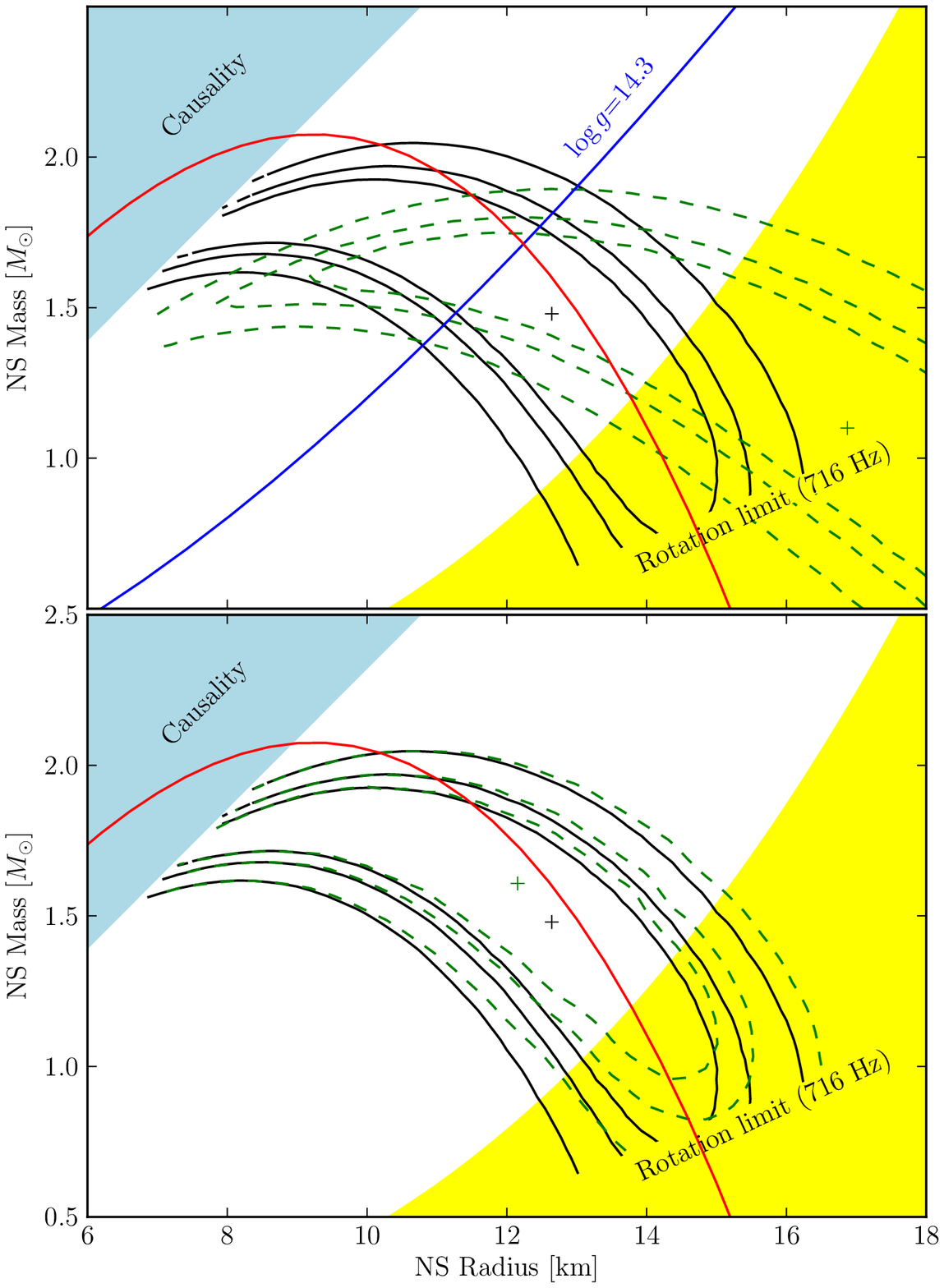}
\caption{\label{f10}
Effect of gravitational acceleration and line-blanketing 
on the best-fit position and confidence contours (50\%, 68\%, and 90\%) 
in the $M$--$R$ plane. The contours are obtained from fitting
an {\sl XMM-Newton}
spectrum of 
the HESS J1731$-$347 CCO with three types of  models, for a fixed distance
$d=3.2$ kpc \citep{Kl:13}.
The solid contours are
obtained using the line-blanketed model atmosphere spectra with 
varying $\log g$ parameter.
The dashed contours in the top panel are obtained 
from a subset of models with fixed $\log g =14.3$. 
The  dashed contours in the bottom panel are obtained 
using
model atmosphere spectra computed without account for spectral lines.
The solid red 
 curves correspond to $R_{\infty}=R(1+z)=16$ km, and the solid blue curve in the top panel
 corresponds to $\log g=14.3$.
}
\end{figure}

Examples of the emergent spectra and temperature structures are shown in
Figures \ref{af1} -- \ref{af3}, while Figure \ref{dens-ioniz} shows the depth dependences
of the density and 
ionization fractions. The absorption edges arising due to 
both CV and CVI ions are seen in the low  
temperature spectra. At higher temperatures, the number density of CV
becomes so small that only the CVI  
edge is seen. The edges modify the spectra dramatically,
causing strong deviations from the  blackbody and hydrogen atmosphere specta
of the same $T_{\rm eff}$
 (see Fig.\,\ref{af2}).   
Since the large opacity above the CVI ionization edge blocks the emergent flux
at the maximum of the Planck function 
$B_{\rm E}(T_{\rm eff})$, this
radiation flux has to be emitted at 
 higher photon energies, where the atmosphere is more transparent.   
As a result,  
a nearly flat spectrum forms  above the energy of
the CVI ionization edge, $E > 0.49$ keV, 
so that the carbon atmosphere spectrum is
harder and has a lower spectral flux density 
 than the hydrogen atmosphere 
spectrum at these energies.
Also, the importance of taking into account the bound-bound transitions (spectral
lines) is evident. The inclusion of the lines modifies the 
spectrum even at photon energies above the edge and leads to 
additional cooling of
the upper atmosphere layers. 
The spectra of higher gravity models are
slightly harder compared to the lower gravity models (see Fig.\ \ref{af3}). 
We note that the absorption from the dissolved atomic levels is 
more important
for the higher gravity models -- see the continuum flux levels between the
Ly$\alpha$- and Ly$\beta$-like spectral lines near  $E \approx$~0.4\,keV. 

In Figure \ref{af4},
the emergent spectra and temperature structures of models with
carbon-hydrogen mix are presented.
It shows that the emergent spectra become very
similar to the pure carbon atmosphere spectra already at
a relatively low carbon 
abundance, $\sim 10\%$ of the hydrogen number density abundance. 

Due to high densities in NS atmospheres, electron
degeneracy may become important at a certain depth. We do not include this 
 effect in our calculations. We can, however, 
evaluate its importance as follows.
For a few models we have calculated the Fermi temperature 
$T_{\rm F} = E_{\rm F} / k$ corresponding to  
the electron number density $N_{\rm e}$: $E_{\rm F} = m_{\rm e} c^2 (\sqrt{1+x^2}-1)$, where 
$x={p_{\rm F}}/{m_{\rm e}c}$, and  $p_{\rm F}=h\left({3}N_{\rm e}/{8\pi}\right)^{1/3}.$
Electron degeneracy becomes important if the Fermi temperature is
comparable to or larger than the gas temperature. 
For the computed models, the spectrum formation depths turn out to be
considerably smaller than the depth 
where the Fermi temperature exceeds the electron gas temperature (see
Fig.\ \ref{af5}). The  spectrum formation depth corresponds to column
density $m$ where emergent photons are created or, equivalently,
$\tau_\nu^{\rm eff}$ = 1.

Another effect of very high densities and/or low temperatures is a
deviation of the equation of state from the ideal gas law due to 
the interactions between particles. Using calculations by Potekhin (1996),
we have checked that such deviations are small in the layers where the
emergent spectrum is formed, at the effective temperatures of our
atmosphere models.  
Therefore, we conclude that neither the electron degeneracy 
nor the deviations from the ideal gas law 
affect the emergent spectra considerably \citep[see also][in particular Fig.\ 6 of that paper]{sw:07}.

We also investigated the accuracy of the method we used to account for 
the pressure 
ionization and level dissolution. 
For this purpose, we computed two
additional model atmospheres where these effects are accounted 
for using different approaches. The first model is
computed in the simplest approach, 
in which the ionization
potential decrease is calculated as due to the Debye screening \citep[see details
in][]{Kurucz:70}.
The second one is computed with the occupation
probabilities $w_{ijk}$ calculated using the 
OP approximation \citep{S94} instead of  the \citet{Lanz.Hub:94} approximation which we used.
 Here, $w_{ijk}$ is the probability that the bound 
level $i$ of the ionization stage $j$ of the chemical
element $k$  is indeed bound. The value (1-$w_{ijk}$) gives the 
probability that this level is dissolved. It 
turns out that the models computed using these two other approaches give
spectra and temperature structures 
very close to those of our model (see Fig.\ \ref{af5a}). This means
that the differences in the pressure ionization and level dissolution 
descriptions 
may become important only at depths larger than the depth of
the spectrum formation.

It is interesting to compare our model spectra
 with those published by
 \citet{HoH:09}  using similar input
 parameters. The result of such a comparison,
shown in Figure \ref{af6}, 
demonstrates a good agreement.

\section{Fitting NS spectra with the carbon atmosphere models}

For fitting observed X-ray spectra with our models, we computed a suite
of model spectra that can be directly used by the XSPEC 
package\footnote{http://heasarc.gsfc.nasa.gov/xanadu/xspec/models/carbatm.html} (XSPEC model \texttt{carbatm}). We converted the 549 
computed model flux spectra  to photon spectra in
the energy range 0--20 keV,
computing the numbers of photons in
1000 equal energy bins of 0.02 keV width.
The fitting parameters of the
model are the effective temperature $T_{\rm  eff}$, the NS mass $M$
(in units of $M_\odot$) and
radius $R$ (in units of km),
and the normalization parameter
$K$. 
For a given set of the fitting parameters, the surface gravity $g$ is
calculated using
Equation (\ref{eq:g_def}), 
the trial spectrum is computed using a linear
interpolation between the nearest model spectra on the
$T_{\rm  eff}$--$\log g$ grid,
and the boundaries of the energy bins and the number of 
photons in each bin are divided  by $(1+z)$, defined by Equation (\ref{eq:redshift_def}). The normalization
is defined as $K=A/d_{10}^2$,
where $d_{10}$ is the distance to the
source in units of 10\,kpc, and $A$  
characterizes the fraction of the NS surface
which  emits X-rays ($A=1$ if the radiation is produced by the entire NS surface).
Obviously, $R$ and $d$ cannot be independently determined from a spectral fit
because the observed flux is proportional to $R^2/d^2$. If the goal is to constrain
the NS radius (and mass) for a source with a well-measured distance, and there
are reasons to believe that the NS surface is uniformly heated ($A=1$),
 then $d$ (and $K$) can be fixed. If the distance is poorly known, then one can fix $R$
at a reasonable value (or set 
reasonable boundaries for varying $R$) and constrain $d$
(or $A^{-1/2}d$) from the fit.

The \texttt{carbatm} model has been tested by \citet{Kl:13}
 on the CCO in the  
HESS J1731$-$347 supernova remnant.
Figure 4 of that paper shows confidence contours in the $M$--$R$ plane
 for two fixed distances and $A=1$.
In Figure \ref{f10} of our paper 
we reproduce those contours for $d=3.2$\,kpc (solid curves).
In the top panel of Figure \ref{f10} we also show confidence contours
(dashed curves) obtained for the same $d$
using the models with fixed $\log g$ = 14.3 (i.e., for any point in the 
$M$--$R$ plane we fit the observed spectrum using the subset of models
 with the fixed 
$\log g$ but applying the correct gravitational redshift). These contours
differ substantially from the correct ones (except for the points close
to the curve $\log g = 14.3$),
which demonstrates the effect of graviational acceleration and
 importance of using models with correct $\log g$ values.
In the bottom panel of Figure \ref{f10} we show the confidence contours
obtained from fitting the models in which bound-bound transitions (spectral
lines) were neglected (dashed curves), which demonstrate the effect of
line blanketing.

\section{Discussion and Summary}

In this work we
have presented a detailed description of our 
carbon atmosphere models and demonstrated how the models can be applied
to constrain the NS parameters.
The models have already been  
used to fit the X-ray spectra of two CCOs, in the HESS\,J1731$-$347
and in Cas~A supernova remnants.
Although the inferred constraints on the NS mass and radius are consistent 
with many equations of state, the fits demonstrate that these NSs may be
indeed covered by carbon atmospheres and radiate from the entire uniformly
heated surface.
 These sources are
good 
potential  
targets  for further detailed investigations of 
the NS properties,
including the NS cooling rates (see \citealt{HHo:10} and \citealt{PPS13}),
with the next  generation of X-ray observatories.

In our work, we have
thoroughly 
analyzed the accuracy of 
our assumptions for the carbon  
atmosphere computations. We 
have checked that our calculations of the carbon opacities and
the number densities of 
carbon ions are consistent with the
results of the Opacity Project. 
We found that 
the thermal conductivity of degenerate electrons, which determines the
energy transport in the NS crust and deep 
atmosphere layers,  becomes insignificant in the layers where
  the spectra are formed. 
We also showed that the model spectrum
computed by \citet{HoH:09} is in close agreement
with the one computed
here.

There are, however, at least two
physical effects 
 which can impact on the accuracy of the presented model spectra.
The first one is the assumption of LTE. We have computed the number
densities of the carbon ions and the population of 
ion excited levels using the Saha-Boltzmann equations with the pressure
ionization effects taken into account.  
In general, these 
quantities are determined by  kinetic
equilibrium between the rates of the radiative and 
collisional transitions 
that populate and de-populate 
ion states. 
Solving these kinetic equations constitutes the 
non-LTE approach \citep[see e.g. ][]{Mihalas:78}. A deviation from the
LTE number density for H- and He-like carbon 
ions can modify the observed part of the
emergent model spectra. The possible importance of this effect has been 
demonstrated for solar abundance NS model atmospheres \citep{RSW08}.

The second effect is the influence
of a magnetic field. The investigated CCOs in Cas~A and HESS J1731$-$347 do not show any
evidence for a magnetic field, but even a relatively weak field
strength ($B < 10^{10}$ G) can 
affect the emergent model spectra
via shifting
the ion energy levels and ionization potentials, 
which changes the ion number densities 
and  positions of the photoionization edges
and spectral lines in emergent 
spectra \citep{PP:95, PP:97, MH:07, EKW:09}
Moreover, even a weak magnetic field
can broaden the carbon spectral lines due to Zeeman
effect increasing the line-blanketing effect, which is 
important even without magnetic field.

 All these effects can be important for
  measuring 
fundamental NS parameters, including masses and radii, and they will
be investigated in future works.

\acknowledgments
The work  of VS is supported by the German Research Foundation
(DFG) grant SFB/Transregio 7 "Gravitational Wave Astronomy" and the 
Russian Foundation for Basic Research (grant 12-02-97006-r-povolzhe-a). 
The work of GGP was partially supported by NASA ADP grant NNX09AC84G
and {\sl Chandra} grant GO2-13083X.

\end{document}